\RequirePackage[l2tabu, orthodox]{nag}
\RequirePackage{snapshot}

\documentclass[9pt,onecolumn]{extarticle}

\makeatletter
\if@twocolumn
  \usepackage[dvips,letterpaper,top=0.5in, bottom=0.5in, left=0.75in, right=0.5in,includefoot,heightrounded]{geometry}
\else
  \usepackage[dvips,letterpaper,margin=1in,includefoot,heightrounded]{geometry}
\fi

\usepackage{srcltx}

\usepackage[russian,portuges,english]{babel}

\iflanguage{portuges}
    {\newcommand{\keywordname}{Palavras-chaves}}
    {\newcommand{\keywordname}{Keywords}}

\usepackage{amsmath}
\usepackage{amssymb,amsfonts}

\usepackage{abstract}

\usepackage{graphicx}
\usepackage[usenames,dvipsnames,svgnames,x11names]{xcolor}
\usepackage{subfigure}

\usepackage{booktabs}

\usepackage{setspace}
\usepackage{flushend}

\usepackage{cite}

\usepackage{hyperref}
\usepackage[normalem]{ulem}
\usepackage{bm}

\usepackage{enumerate}

\usepackage{multirow}

\usepackage{algorithm}
\usepackage{algorithmic}

\usepackage{listings}

\usepackage[short,12hr]{datetime}
       \usepackage{fouriernc}

\newcommand{\diag}{\operatorname{diag}}

\makeatletter

\newcommand{\printtitle}{%
\makeatletter
\if@twocolumn

\twocolumn[%
  \maketitle
  \begin{onecolabstract}
    \myabstract
  \end{onecolabstract}
  \begin{center}
    \small
    \textbf{\keywordname}
    \\\medskip
    \mykeywords
  \end{center}
  \bigskip
]
\saythanks
\else
  \maketitle
  \begin{onecolabstract}
    \myabstract
  \end{onecolabstract}
  \begin{center}
    \small
    \textbf{\keywordname}
    \\\medskip
    \mykeywords
  \end{center}
  \bigskip
  \onehalfspacing
\fi
\makeatother
}

\author{
B. G. Palm%
	\thanks{Department of
		Telecommunications,
		Aeronautics Institute
		of Technology (ITA),
		Brazil
		and
		Department of
		Mathematics
		and Natural Sciences,
		Blekinge Institute of Technology,
		Sweden
		(E-mail: \protect\url{brunagpalm@gmail.com}).
	}
\and
F. M. Bayer%
	\thanks{Departamento de Estat\'istica
		and LACESM,
		Universidade Federal de Santa Maria, Brazil
		(E-mail: \protect\url{bayer@ufsm.br}).}
\and
R. Machado%
	\thanks{Department of
		Telecommunications,
		Aeronautics Institute
		of Technology (ITA),
		Brazil
		(E-mail: \protect\url{rmachado@ita.br}).
	}
\and
M. I.Pettersson%
\thanks{
Department of
	Mathematics
	and Natural Sciences,
	Blekinge Institute of Technology,
	Sweden
	(E-mail: \protect\url{mats.pettersson,viet.thuy.vu@bth.se}).
}
\and
V. T. Vu${}^\mathsection$%
\and
R. J. Cintra%
\thanks{Signal Processing Group,
	Departamento de Estat\'{\i}stica,
	Universidade Federal Pernambuco, Brazil
	(E-mail: \protect\url{rjdsc@de.ufpe.br}).}
}

\title{%
Robust Rayleigh Regression Method for SAR Image Processing in Presence of 	Outliers
}

\newcommand{\myabstract}{%
The presence of
outliers~(anomalous values)
in synthetic aperture radar~(SAR) data and
the misspecification in statistical image models
may result in inaccurate inferences.
To avoid such issues,
the Rayleigh regression model based on
a robust estimation process
is proposed  as
a more realistic approach to model this type of data.
This paper aims at obtaining
Rayleigh regression model
parameter estimators
robust to the presence of outliers.
The proposed approach considered
the weighted maximum likelihood method
and was submitted
to numerical experiments using
simulated and measured SAR images.
Monte Carlo simulations were
employed
for the numerical assessment
of the proposed
robust estimator performance
in finite signal lengths,
their sensitivity to outliers,
and the breakdown point.
For instance,
the non-robust estimators
show a relative bias value
$65$-fold larger than the results provided by the
robust approach
in corrupted signals.
In terms of
sensitivity analysis and break down point,
the robust scheme
resulted in a reduction
of about~$96\%$
and~$10\%$,
respectively,
in
the mean absolute value of both measures,
in compassion to the non-robust estimators.
Moreover, two SAR data sets
were used to compare the ground type
and anomaly detection results
of the proposed robust scheme
with competing methods in the literature.
}

\newcommand{\mykeywords}{%
Outliers,
Rayleigh regression model,
Robust estimation,
SAR images
}

\date{}

\begin{document}

\printtitle

Synthetic aperture radar~(SAR) data
plays an important role
in remote sensing applications~\cite{oliver2004}
due to its capability
of providing
(i)~wide terrain coverage in a
short observation
of
time and
(ii)~suitable visual information acquisition,
independent of
weather and illumination conditions~\cite{melamed2012anomaly}.
However,
SAR images
are frequently
contaminated
by a percentage
of
outliers~(anomalous
values)---image pixels
that differ significantly
from their neighborhood~\cite{grubbs1969}---,
which can be related to
human-made objects
or
highly
reflective areas~\cite{bustos2002m}.
Because such
observations
do not follow
the
general behavior
of the neighborhood
or the observed scene~\cite{melamed2012anomaly},
the use of
suitable approaches to deal with outliers
should be considered
to avoid unreliable results
in remote sensing applications~\cite{bustos2002m}.
For example,
in~\cite{Palm2020},
the median was
employed to obtain a ground scene predicted~(GSP)
image
based on an image stack of the CARABAS~II data set~\cite{Lundberg2006}.
The resulted image was applied
as a reference image in a change detection algorithm.
The median is a robust method and was
considered in the CARABAS II SAR image data set
to remove outliers~(military vehicles),
resulting in high probability detection values,
associated with low false alarm rates,
highlighting the importance
of robust methods to deal with outliers.

Typical tasks
in
SAR data analysis and processing
include
(i)~image modeling~\cite{Bustos2009a,sportouche2017},
(ii)~identification
and
classification
of
distinct
ground type~\cite{
	tison2004,
	inglada2007},
and
(iii)~change detection~\cite{mercier2008,zhao2018robust}.
The use of
statistical models---commonly employed to
describe image pixels by
a small number of parameters~\cite{allende2001robust}---can
generate accurate results for the above SAR-related challenges,
as presented in~\cite{torres2017} and~\cite{wang2008}.
The statistical inference methods
widely considered
for signal and image modeling
usually suppose
(i)~Gaussian
or
symmetric data~\cite{allende2001robust,zoubir2018robust}
and
(ii)~least-squares or maximum likelihood approaches~\cite{allende2001robust}.

However,
magnitude
SAR image
pixels
generally
present non-Gaussian
properties,
such as
asymmetrical distributions
and
strictly
positive
values~\cite{oliver2004}.
These characteristics
motivated the proposition of
a regression model
based on the Rayleigh distribution
for SAR image modeling~\cite{Palm2019}.
The  Rayleigh regression model
is suitable for
non-Gaussian situations,
where the observed output signal is asymmetric
and measured continuously on the real positives values,
such as SAR amplitude image pixels.
The Rayleigh regression model
assumes
that the Rayleigh distributed signal mean follows
a regression structure involving covariates,
unknown parameters, and a link function.
In~\cite{Palm2019},
an inference approach for the model parameters,
diagnostic tools,
asymptotic proprieties of the parameter estimators,
and a ground type detector
were discussed.
Additionally,
as in other classes
of non-Gaussian regression models,
such as
the generalized linear model~\cite{McCullagh1989},
the Rayleigh regression model
was derived
considering the maximum likelihood
approach
to estimate its parameters.
However,
robust tools were not discussed.

The maximum likelihood inference method
is asymptotically efficient but lacks
robustness against
model misspecification
and outliers~\cite{ghosh2016robust,zoubir2018robust}.
On the other hand,
robust approaches
are not significantly affected
by outliers or small model departures~\cite{zoubir2018robust}.
Consequently,
aiming at avoiding
corrupted results
related to the presence of outliers
in the signal of interest,
robust approaches for
the ordinary linear regression models
and
generalized linear models
have been discussed;
see, e.g.,~\cite{ghosh2016robust}
and~\cite{ghosh2013robust}.
Additionally,
as discussed in~\cite{susaki2004robust},
robustness
is an
important
feature
to obtain meaningful physical
estimated parameters in remotely sensed data,
since
robust statistical processing
involves making inferences
in
distorted or corrupted signals~\cite{zoubir2018robust,zoubir2014},
such as SAR data.
In~\cite{Palm2020},
the median was applied in an image stack
to obtain a ground scene predicted image,
while in~\cite{Palm2019},
the Rayleigh regression model was introduced;
both schemes were considered in
remote sensing applications.
In particular, the GSP image was employed
in a change detection algorithm,
and the Rayleigh regression model was considered
in a ground type detection tool.
However,
to the best of our knowledge,
a robust approach
for the Rayleigh regression model
parameter estimation
is absent in the literature,
and this paper aims at proposing the first treatment.
In this paper, our goal is twofold.
First, we derive
a robust statistical tool
for the Rayleigh regression model
for corrupted signals.
Specifically,
to obtain the parameter estimators
robust to the presence of outliers,
we employed the
weighted maximum likelihood method~\cite{field1994robust}.
We introduce parameter estimation
and
large data record inference.
Monte Carlo simulations are used to evaluate the
finite signal length performance
of the Rayleigh regression model robust parameter estimators,
its sensitivity to outliers,
and
the breakdown point.
Second,
this paper
attempts to establish
a framework for detection tools
in SAR images corrupted with outliers
according to the following methodology.
\begin{enumerate}
\item We use the proposed
robust
approach
to detect
ground types
in
the
magnitude single-look
SAR images
obtained from:
\begin{itemize}
	\item CARABAS~II,
	a Swedish
	ultrawideband (UWB)
	very-high frequency~(VHF)
	SAR
	system;
	\item OrbiSAR,
	a Brazilian~SAR system operating at~ X- and~P-bands.
\end{itemize}
\item We employ
the introduced robust scheme
to detect
targets in a CARABAS~II SAR image,
since this
data set is
widely explored in
the literature for
detection of military vehicles concealed by forest;
see, e.g.,~\cite{Palm2020},~\cite{Lundberg2006},~\cite{Ulander2005},
\cite{Vu2017}, and~\cite{Vu2018}.
Such targets
can be interpreted as
anomalies,
since they
introduce more
representative
behavior changes
in the
CARABAS~II
ground scene.
\end{enumerate}

The paper is organized as follows.
Section~\ref{s:model}
reviews the Rayleigh regression model,
introduces robust parameter
estimation and large data record properties.
Section~\ref{s:num}
shows
a
Monte Carlo
simulation
study
for numerical evaluation
of the introduced approach,
using
breakdown point
and sensitivity analysis.
Section~\ref{s:appli}
displays
two experiments
with two measured SAR data sets.
Finally, the conclusion of this work
can be found in Section~\ref{s:con}.

\section{The Proposed Robust Rayleigh
	Regression Method}
\label{s:model}

This section introduces
a robust estimation approach
for the Rayleigh regression model parameters
based on the weighted maximum likelihood method.
Moreover, large data record inferences are discussed.

\subsection{The Rayleigh Regression Model}
\label{s:rrmodel}

\begin{figure}
	\centering
	\includegraphics[width=0.5\textwidth]{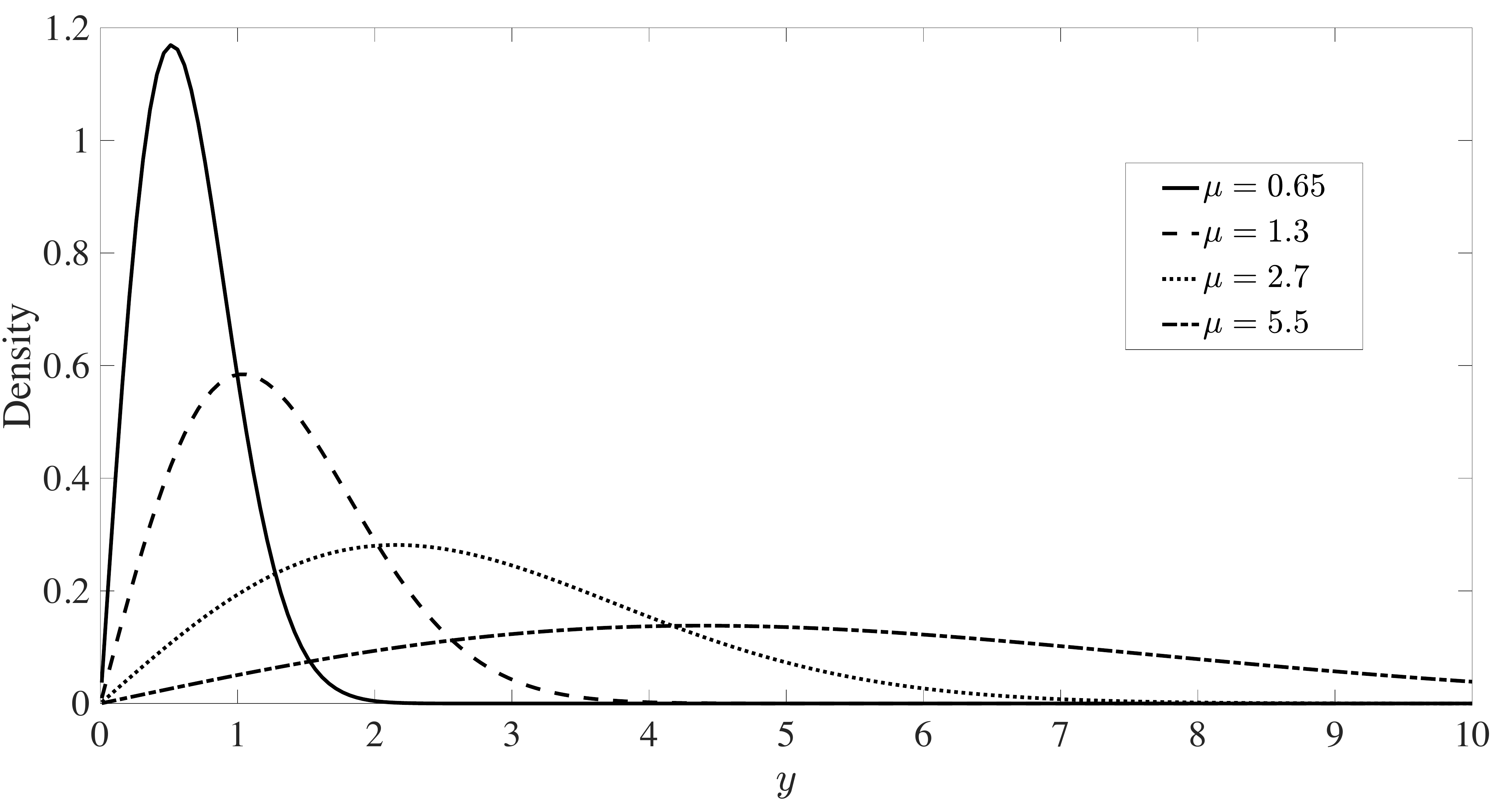}
	\caption{
		Rayleigh probability densities functions
		for~$\mu \in \lbrace 0.65, 1.3, 2.7, 5.5 \rbrace$.
	}\label{f:fig}
\end{figure}

The Rayleigh regression model
introduced in~\cite{Palm2019}
can be defined as follows.
Let~$Y[n]$,~$n = 1,2,\ldots , N$,
be
a
Rayleigh
distributed
random
variable
and let~$y[n]$
be
the realization
of the signal~$Y[n]$
with mean~$\mu[n]$.
Considering the
mean-based parametrization
of~$Y[n]$,
we have that
the probability density function
of~$Y[n]$
is written as
\begin{align}
\label{e:den}
f_Y(y[n];\mu[n] )
=
\frac{\pi y }{2 \mu[n]  ^2}
\exp\left(-\frac{\pi y[n] ^2}{4 \mu[n] ^2}\right)
,
\end{align}
where~$\operatorname{E}(Y[n])=\mu[n] > 0$.
Also,
we have that~$\operatorname{Var}(Y[n])= \mu[n] ^2 \left(\frac{4}{\pi}-1 \right)$.
Figure~\ref{f:fig}
shows
a few different
Rayleigh densities
along with
the corresponding mean parameter~($\mu$) values.
It is noteworthy that the Rayleigh distribution
is flexible,
displaying
different shapes depending on the mean parameter value.
The cumulative distribution function
and the
quantile function
are provided, respectively, by
\begin{align}
F_Y(y[n];\mu[n])
=&
1- \exp\left(-\frac{\pi y[n]^2}{4 \mu[n]^2}\right)
,
\\
Q_Y(u[n];\mu[n])
=&
2 \mu [n] \sqrt{ \frac{-\log(1-u[n])}{\pi} }
.
\end{align}
The quantile
function is
useful for generating
non-uniform
pseudo-random occurrences
according to
the inversion
method---probability
integral transform (PIT)---,
which involves
computing the quantile function
and then inverting it~\cite[Chapter 2]{devroye1986}.
The cumulative distribution function
is employed to define the quantile residuals~\cite{Dunn1996},
which is derived based on~$F_Y(y[n];\mu[n])$
and on standard normal quantile function.
Both methods
are considered in
the simulation and SAR image studies
presented in this paper.

The Rayleigh regression model is
defined
assuming that the mean~$\mu[n]$
of
the observed output signal~$Y[n]$
can be written as
\begin{align}
\label{e:model}
\eta[n]
=
g(\mu[n]) =
\sum_{i=1}^{k}  \beta_i  x_{i}[n]
,
\quad
n=1,2,\ldots, N,
\end{align}
where~$k<N$ is the
number
of covariates considered
in the model,
$\bm{\beta} = (\beta_1, \beta_2, \ldots, \beta_k)^{\top}$
is the
vector
of unknown
parameters,
$\mathbf{x}[n]=(x_{1}[n], x_{2}[n], \ldots, x_{k}[n])^\top$
is the
vector
of
deterministic
independent
input
variables,~$g:\mathbb{R}^+\!\rightarrow\mathbb{R} $
is a strictly monotonic and twice
differentiable link function,
and~$\eta[n]$
is the
linear predictor.
Parameter estimation
based on the maximum likelihood method,
diagnostic measures,
and
further mathematical properties,
including large data record results,
are fully discussed in~\cite{Palm2019}.

\subsection{Robust Estimation}

Robust parameter estimation of the
Rayleigh
regression model can be performed
by
the weighted maximum likelihood approach~\cite{field1994robust}.
Given a
known
weighted
vector~$\mathbf{w}
= (w[1], w[2], \ldots , w[N])^\top
$,
the weighted maximum likelihood
estimates are given by
\begin{align}
\widehat{\bm{\beta}}
=
\arg
\max_{\bm{\beta}}
\ell_w(\bm{\beta})
,
\end{align}
where~$\ell_w(\bm{\beta})$ is the
weighted log-likelihood
function
of the parameters
for the observed signal,
defined as
\begin{align}
\ell_w(\bm{\beta})=\sum_{n=1}^{N}
w[n]
\ell[n](\mu[n])
.
\end{align}
The quantity~$\ell[n](\mu[n])$
is
the logarithm of~$f(y[n],\mu[n])$,
which is
written as
\begin{align}
\begin{split}
\ell[n](\mu[n])
=&
\log\left(\frac{\pi}{2}\right) +
\log(y[n])
\\&
-
\log(\mu[n]^2)-\frac{\pi y[n]^2}{4 \mu[n]^2}
,
\end{split}
\end{align}
where~$\mu[n]= g^{-1}\left(\sum_{i=1}^{k} x_{i}[n] \beta_i \right)$.
The weighted score vector,
obtained by differentiating the
weighted
log-likelihood function
with respect to each unknown parameters~$\beta_i$,
$i = 1,2, \ldots, k$,
is given by
\begin{align}
U_w(\bm{\beta})=
\left(
\frac{\partial \ell_w(\bm{\beta})}{\partial \beta_1},
\frac{\partial \ell_w(\bm{\beta})}{\partial \beta_2},
\ldots,
\frac{\partial \ell_w(\bm{\beta})}{\partial \beta_k}
\right)^\top
.
\end{align}
Considering the chain rule,
we have that
\begin{align}
\frac{\partial \ell_w(\bm{\beta})}{\partial \bm{\beta}}=
\sum_{n=1}^{N}
w[n]
\frac{d \ell[n](\mu[n])}{d \mu[n]}
\frac{d \mu[n]}{d \eta[n]}
\frac{\partial \eta[n]}{\partial \bm{\beta}}
.
\end{align}
As reported in~\cite{Palm2019},
note that
\begin{align}
\label{e:deri}
\frac{d \ell[n](\mu[n])}{d \mu[n]}
&= \frac{\pi y[n]^2}{2\mu[n]^3}-\frac{2}{\mu[n]}
,
\\
\frac{d \mu[n]}{ d \eta[n]}
&= \frac{1}{g'(\mu[n])},
\\
\frac{\partial \eta[n]}{\partial \beta_i}
&= x_{i}[n]
,
\end{align}
where~$g'(\cdot)$ is the first derivative of
the adopted link function~$g(\cdot)$.
In matrix form,
the weighted score vector can be written as
\begin{align}
U_w(\bm{\beta})
=
\mathbf{X}^\top
\cdot
\mathbf{W}
\cdot
\mathbf{T}
\cdot
\mathbf{v}
,
\end{align}
where
$\mathbf{X}$ is an~$N \times k$
matrix whose~$n$th row
is~$\mathbf{x}[n]^\top$, and
\begin{align*}
\mathbf{W}
=&
\diag
\lbrace
w[1], w[2], \ldots , w[N]
\rbrace
,
\\
\mathbf{T}
=&
\diag\left\{
\frac{1}{g'(\mu[1])},
\frac{1}{g'(\mu[2])},
\ldots,
\frac{1}{g'(\mu[N])}
\right\}
,
\\
\mathbf{v}
\!
=&
\!\left(\!\frac{\pi y[1]^2}{2 \mu[1]^3}
\!
-
\!
\frac{2}{\mu[1]},
\frac{\pi y[2]^2}{2 \mu[2]^3}
\!
-
\!
\frac{2}{\mu[2]},
\ldots,
\frac{\pi y[N]^2}{2 \mu[N]^3}
\!
-
\!
\frac{2}{\mu[N]} \!\right)\!^\top\!\!
.
\end{align*}
Thus,
the weighted maximum likelihood estimator~(WMLE)
for each
Rayleigh
regression model parameter is obtained
by solving the
following nonlinear system
\begin{align}
\label{e:vescore}
U_w(\bm{\beta}) = \mathbf{0}
,
\end{align}
where~$\mathbf{0}$ is the~$k$-dimensional vector of zeros.
The
quasi-Newton Broyden-Fletcher-Goldfarb-Shanno
(BFGS) method~\cite{press}
with analytic first derivatives
was considered as the nonlinear optimization algorithm
to solve~\eqref{e:vescore}.
To determine the initial points,
we followed the same methodology
described in~\cite{Palm2019}.

\subsubsection{Weighted Vector Determination}

The WMLE is obtained supposing that the weights $w[n]$,
$n=1,2,\ldots,N$, are known.
However,
in practice,
we have to determine these values.
As suggested by \cite{field1994robust},
we consider the following approach for the
weighted vector determination:
\begin{align}
\label{e:pesos}
w[n] =
\begin{cases}
\frac{F(y[n]; \mu[n])}{\delta}
,
& \text{if} \, \, F(y[n]; \mu[n]) < \delta
,
\\
1,
& \text{if} \, \, \delta \leq F(y[n]; \mu[n]) \leq 1-\delta
,
\\
\frac{1-F(y[n]; \mu[n])}{\delta},
& \text{if} \, \, F(y[n]; \mu[n]) > 1-\delta
,
\end{cases}
\end{align}
where~$\delta \in (0,1)$
is employed to
delimit the weighed distribution interval
in~$(1 - 2 \delta ) \%$
and
the unknown parameter
is replaced by their non-robust
maximum likelihood estimator~(MLE).
Typical values for~$\delta$
are $0.01$ and $0.001$~\cite{field1994robust}.
We note that
atypical~$y[n]$ values imply
in
large or small
$F(y[n]; \mu[n])$ values,
which
are
weighted;
consequently,
inference distortions
related to these observations
are
minimized.

\subsection{Testing Inference}
\label{s:fisher}

Under the following
mild regularity conditions:
(i)~the
first- and
second-order derivatives of the
weighted
log-likelihood function are well-defined;
and
(ii)~the
expectation of the score function is equal to zero,
it is shown
in~\cite{field1994robust} and~\cite{hu2002weighted}
that
the WMLEs
are
asymptotically equivalent to the MLEs.
Thus, we can use the
classical Wald statistic
to make
inferences about
the regression parameters.
Suppose that
we partition the
parameter
vector~$\bm{\beta}$
into a
vector of parameters of interest~($\bm{\beta}_I$),
with dimension~$\nu$,
and
in a
nuisance parameter vector~($\bm{\beta}_M$),
with dimension~$k-\nu$.
The interest hypothesis~$\mathcal{H}_0$
and
the alternative hypothesis~$\mathcal{H}_1$
are given by
\begin{align}
\begin{cases}
\mathcal{H}_0:
\bm{\beta}_{I}=\bm{\beta}_{I0}
,
\\
\mathcal{H}_1:
\bm{\beta}_{I} \neq \bm{\beta}_{I0}
,
\end{cases}
\end{align}
where~$\bm{\beta}_{I0}$ is a fixed column vector of dimension~$\nu$.
The
Wald statistic is given
by~\cite{Kay1998-2}
\begin{align}
T_W = (\widehat{\bm{\beta}}_{I1}-\bm{\beta}_{I0})^\top
\left( \left[ \mathbf{I} ^ {-1}
(\widehat{\bm{\beta}}_1) \right] _{\beta_I \beta _I} \right)^{-1}
(\widehat{\bm{\beta}}_{I1}-\bm{\beta}_{I0}),
\end{align}
where
$\widehat{\bm{\beta}}_1 = (\widehat{\bm{\beta}}_{I1}^ \top ,
\widehat{\bm{\beta}}_{M1}^\top)^\top$
is the
WMLE
under~$\mathcal{H}_1$,
$\mathbf{I}(\widehat{\bm{\beta}})$ is the Fisher information matrix derived in~\cite{Palm2019}
evaluated at the WMLE,
and
$
\left[ \mathbf{I}^{-1} (\widehat{\bm{\beta}}) \right]_{\beta_I \beta _I}
$
is a
partition
of~$\mathbf{I}(\widehat{\bm{\beta}})$
limited to the
interest estimates.
The
Fisher information matrix
is given by
$
\mathbf{I}(\bm{\beta})=
\mathbf{X}^\top
\cdot
\mathbf{W}
\cdot
\mathbf{X}
$,
where
\begin{align}
\begin{split}
\mathbf{W}
&=
\diag\left\{
\frac{4}{\mu[1]^2}\left(\dfrac{d\mu[1]}{d\eta[1]}\right)^2,
\frac{4}{\mu[2]^2}\left(\dfrac{d\mu[2]}{d\eta[2]}\right)^2,
\ldots,
 \right.
 \\&
 \left.
\frac{4}{\mu[N]^2}\left(\dfrac{d\mu[N]}{d\eta[N]}\right)^2
\right\}
.
\end{split}
\end{align}
In particular,
for the log link function~($g(\mu[n]) = \log(\mu[n])$),
the Fisher information matrix
is exactly the same
for the robust and non-robust approaches,
since~$\dfrac{d\mu[n]}{d\eta[n]}
= \mu[n]$.

Based on the consistency of the WMLE
and
the asymptotic normality of the estimators,
$T_W$
statistic
follows,
in large data records,
a
chi-square distribution
with~$\nu$
degrees of freedom,
$\chi^2_\nu$.
The test is performed by comparing the calculated
value
of~$T_W$
with
a threshold value,~$\gamma$,
obtained
from
the~$\chi^2_\nu$ distribution
and
the
desired
probability of false alarm~\cite{Kay1998-2}.
The Wald test described above
can be used
for several detection signal applications,
such as
ground type use
and
presence of a signal.

\section{Simulation Study}
\label{s:num}

This section considers
Monte Carlo simulations to evaluate
the finite signal length performance
of the robust point estimators
of the Rayleigh regression model parameters.
For such,
we assess the estimation performance with
and without outliers.
We also
measure the estimator sensitivity
in the presence of anomalous observations
and
the breakdown point.

\subsection{Robust Point Estimators  Performance }

The numerical
evaluation was made over~$5,000$
different signal samples generated
by means of~\eqref{e:model}
and
considering the log
link function.
Following the methodology
described in~\cite{Palm2019},
the parameters
were adopted as follows:
$\beta_1= 0.5$ and~$\beta _2 = 0.15$,
and
the covariate
was obtained
from the uniform
distribution~$(0,1)$,
and considered constants
for all Monte Carlo replications.
In each replication,
the inversion method
was employed to
simulate~$y[n]$
assuming the Rayleigh distribution
with mean
$\mu[n]=\exp\left\{\beta_1 + \beta_2 x_2[n]\right\}$.

The simulation study considered
signals in several situations,
varying the sample size~$N \in \lbrace
100,\,500,\,750
\rbrace$
and the contamination level~$\epsilon \in \lbrace
0\%, 1\%, 5\%
\rbrace$.
The selected values of~$\epsilon$
follow the literature
as shown
in~\cite{allende2003robust},
\cite{Bustos2009a},
and~\cite{Bayer2020}
for
robust estimation analysis in one-, two-,
and three-dimensional models,
respectively.
The outliers
were included, assuming a value equal to~$10$
in randomized
positions.
We employed~$\delta  \in \lbrace 0.001, 0.01 \rbrace$
for the weight determination;
however,
for brevity and
similarity of results,
just the ones for $\delta = 0.001$ are shown.
The
percentage relative bias~(RB\%),
the mean square
error~(MSE),
and the sum of
the absolute values of
RB\% and
MSE~total
were
adopted as figures of merit to numerically evaluate
the proposed robust estimators.
Such error measures
are expected to be
as small as possible
and
were
computed
between~$\bm{\beta}$
and~$\bm{\widehat{\beta}}$.

Table~\ref{t:reg1}
shows the
Monte Carlo simulation
results
for the
point estimators
of the
Rayleigh regression
model parameters
with and without outliers.
Both WMLEs and MLEs
show similar and small values of RB\% and MSE
for the data without outliers.
In particular,
the absolute total value
of relative bias
for $N=100$
is equal to~$1.6228$
and~$1.5765$
for WMLE and MLE,
respectively.
However,
the MLEs present
higher values of RB\% and MSE
for the contaminated data
when compared to the WMLE results,
showing that the robust theory
is effective, reducing
considerably the
RB\% and MSE values
concerning
the non-robust estimation method
in corrupted signals.
For instance,
consider the case with~$5\%$ of contamination
and~$N=500$, the WMLE for~$\beta_1$
displays values of RB\% equal to~$1.8567\%$,
while the MLE shows RB\% value about~$106\%$
for the same parameter.
Summarizing,
the  WMLEs
show either equal or better performance compared
with the results from
the MLEs, in all evaluated cases.

\begin{table*}
	\caption{
		Results of the Monte Carlo simulation
		of the point estimation considering
		the robust~(WMLE) and non-robust~(MLE) approaches,
		with and without outliers,
		for~$\beta_1=0.5$,~$\beta_2=0.15$, and~$\delta=0.001$
	}
	\label{t:reg1}
	\centering
	\begin{tabular}{ccccc|ccc}
		\toprule
		& &  \multicolumn{3}{c|}{WMLE} &   \multicolumn{3}{c}{MLE} \\
		\midrule
		$\epsilon$ & Measures 	 &	$\widehat{\beta}_1$ 	&	$\widehat{\beta}_2$  	&
		Absolute Total & $\widehat{\beta}_1$ 	&
		$\widehat{\beta}_2$ & Absolute Total	\\
		\midrule
		\multicolumn{8}{c}{ $N=100$ } \\
		\midrule
		\multirow{3}{*}{$0 \%$} & Mean & $0.4957$ &    $0.1488$ &    ---   & $0.4967$ &    $0.1486$ &  --- \\
		& RB(\%)   & $-0.8570$ &   $-0.8259$ &    $1.6828$ &   $-0.6598$ &   $-0.9167$ &    $1.5765$ \\
		& MSE   & $0.0100$ &    $0.0312$ &    $0.0413$ &    $0.0099$ &    $0.0309$ &    $0.0409$  \\
		\midrule
		\multirow{3}{*}{$1 \%$} & Mean &  $0.4968$ &    $0.1480$ &    --- &    $0.6401$ &    $0.1247$ &    --- \\
		& RB(\%)   & $-0.6311$ &   $-1.3311$ &    $1.9622$ &  $ 28.0262$ &  $-16.8944$ &   $44.9206$ \\
		& MSE   &  $0.0101$ &    $0.0316$ &    $0.0417$ &    $0.0715$ &    $0.1916$ &    $0.2631$ \\
		\midrule
		\multirow{3}{*}{$5\%$} & Mean &      $0.5476$ &  $0.1597$ &  --- &  $1.0103$ & $0.0705$ & --- \\
		& RB(\%)   &  	$9.5122$ &    $6.4767$ &   $15.9889$ &  $102.0672$ &  $-53.0026$ &  $155.0698$ \\
		& MSE   &  	$0.0518$ &    $0.1693$ &    $0.2211$  &  $0.3405$ &    $0.3098$ &    $0.6503$ \\
		\midrule
		\multicolumn{8}{c}{ $N=500$ } \\
		\midrule
		\multirow{3}{*}{$0 \%$} & Mean & $0.4974$ &    $0.1508$ &    --- &    $0.4989$ &    $0.1508$ & ---    \\
		& RB(\%)   &  $-0.5234$ &    $0.5316$ &    $1.0550$ &   $-0.2291$ &    $0.5216$ &    $0.7507$\\
		& MSE   & $0.0021$ &    $0.0062$ &    $0.0083$ &    $0.0020$ &    $0.0061$ &    $0.0081$ \\
		\midrule
		\multirow{3}{*}{$1\%$} & Mean & $0.4989$ &    $0.1506$ &    ---  &  $0.6579$ &    $0.1183$  & ---  \\
		& RB(\%)   & $-0.2117$ &    $0.4203$ &    $0.6320$ &   $31.5790$ &  $-21.1080$ &   $52.6870$   \\
		& MSE   & $ 0.0021$ &    $0.0062$ &    $0.0083$ &    $0.0362$ &    $0.0425$ &    $0.0786$   \\
		\midrule
		\multirow{3}{*}{$5\%$} & Mean &    $0.5093$  &   $0.1628$ &    --- &    $1.0314$ &    $0.0679$ &    --- \\
		﻿& RB(\%)   &    $1.8567$ &    $8.5400$ &   $10.3967$ &  $106.2793$ &  $-54.7086$ &  $160.9879$ \\
		﻿& MSE   &   $0.0045$ &    $0.0162$ &   $ 0.0208$ &    $0.2954$ &    $0.0566$ &    $0.3519$ \\

		\midrule
		\multicolumn{8}{c}{ $N=750$ } \\
		\midrule
		\multirow{3}{*}{$0 \%$} & Mean & $0.4980$ &    $0.1507$ &    --- &    $0.4996$ &    $0.1506$ &    ---\\
		& RB(\%)   & $-0.3962$ &    $0.4568$ &    $0.8530$ &   $-0.0875$ &    $0.4140$ &    $0.5015$  \\
		& MSE   & $0.0014$ &    $0.0041$ &    $0.0054$ &    $0.0014$ &    $0.0040$ &    $0.0054$  \\
		\midrule
		\multirow{3}{*}{$1\%$} & Mean & $0.4997$ &    $0.1507$ &    ---  &  $0.6673$ &    $0.1184$ &    ---  \\
		& RB(\%)   & $-0.0593$ &    $0.4608$ &    $0.5201$ &   $33.4529$ &  $-21.0987$ &   $54.5516$ \\
		& MSE   & $0.0014$ &    $0.0041$ &    $0.0054$ &    $0.0360$ &    $0.0300$ &    $0.0660$  \\
		\midrule
		\multirow{3}{*}{$5\%$} & Mean &   $0.5080$  &  $0.1631$ &  ---  & $1.0359$ &  $0.0698$ &    --- \\
		﻿& RB(\%)   &   $1.6012$ &    $8.7438$ &   $10.3450$ &  $107.1780$ &  $-53.4431$ &  $160.6211$ \\
		﻿& MSE   &    $0.0027$ &    $0.0099$ &    $0.0127$ &    $0.2959$ &    $ 0.0396$ &    $0.3355$ \\

		\bottomrule
	\end{tabular}
\end{table*}

\subsection{The Breakdown Point and Sensibility Curve}

\begin{figure*}
	\centering
	\subfigure[Total breakdown point]{
		\includegraphics[scale=0.3]{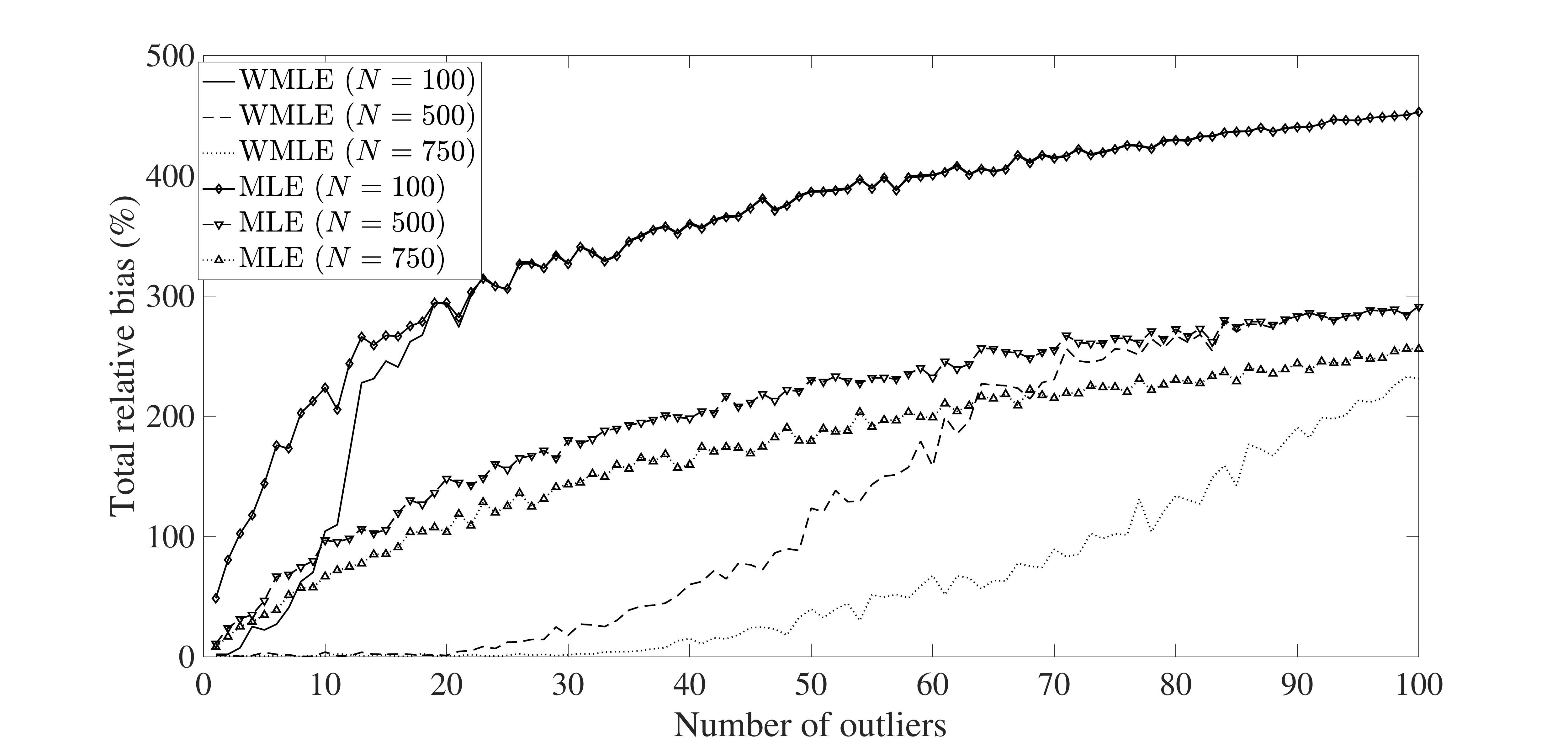}
		\label{f:rup}}
	\subfigure[Sensitivity curves]{
		\includegraphics[scale=0.3]{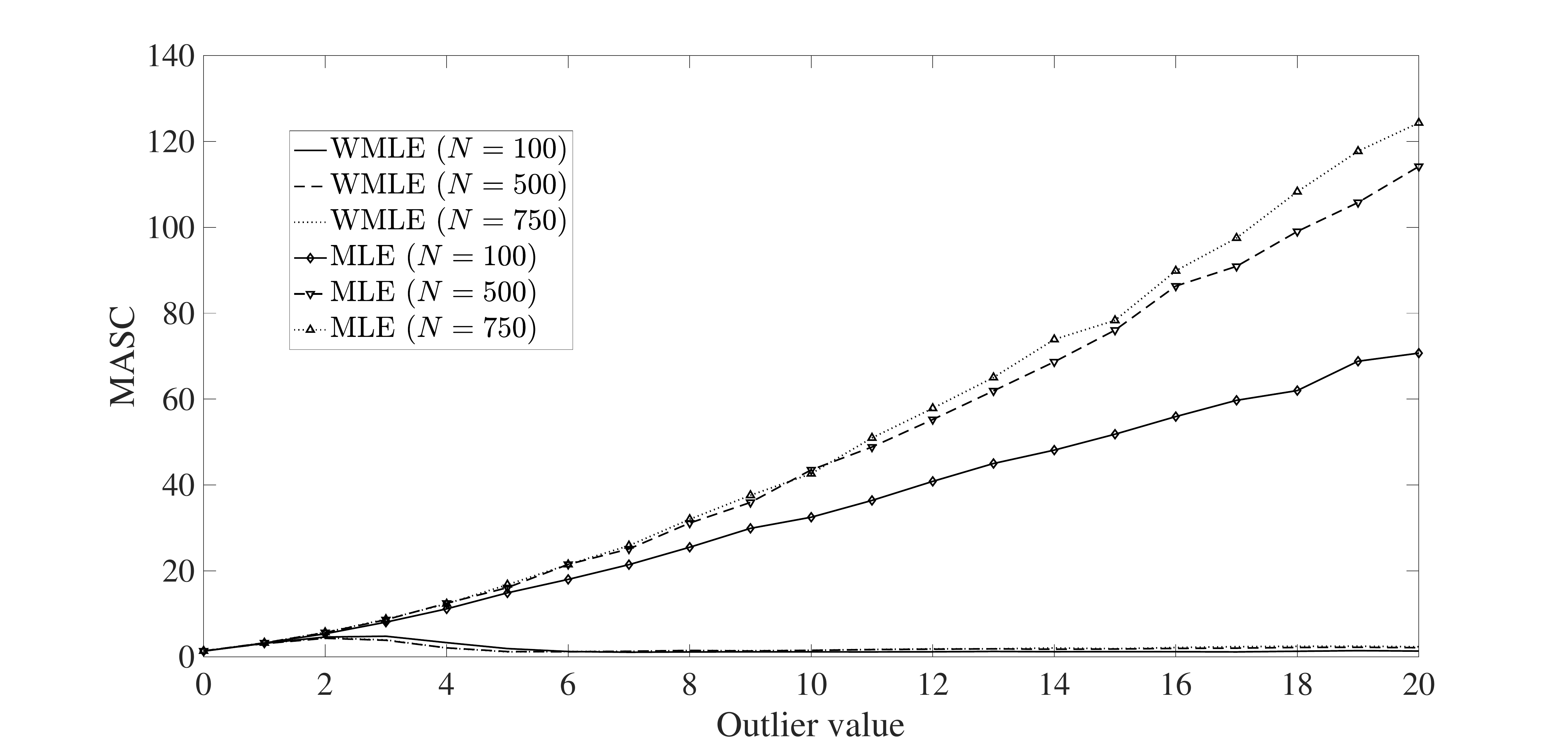}
		\label{f:masc}}
	\caption{The total breakdown point
		and
		sensitivity results considering $5\%$
		of outliers
		and
		$1000$ Monte Carlo replications.
	}

\end{figure*}

The breakdown point
was proposed in~\cite{hampel1971general}
and evaluate the
proportion of outliers that the signal
may contain such that~$\widehat{\bm{\beta}}$
still provides some information about the true parameter~\cite{hampel1971general,yohai1987high,zoubir2018robust}.
Figure~\ref{f:rup}
displays the total
breakdown point
in terms of the
total relative bias
of the estimators,
which is defined
as the sum of the absolute values of the individual relative biases.
As in the  robust estimators
point evaluation,
the outlier value was set equal to~$10$.
Additionally,
we employed
$1000$ Monte Carlo replications
and
the number of outliers ranging from $1$ to $100$.
We note that the WMLEs show
smaller total relative bias values
compared with the results from the MLEs,
in all evaluated cases.
In general,
we note that
for~$1\%$ of contamination,
the MLEs present total relative bias close to or
higher than $100$.
On the other hand,
the WMLEs
show the same total relative bias values
for a
contamination level of about~$10\%$
of the observations.

Another measure widely used
to evaluate robust estimators is the
sensitivity curve (SC) \cite{zoubir2018robust},
which
provides
an
intuitive information
about the sensitivity of an estimator
measuring its variability
with the addition of an outlier to the signal.
The SC is given by~\cite{zoubir2018robust}
\begin{align}
\begin{split}
\text{SC} (\mathbf{y}, \bm{\widehat{\beta}})
&= N \cdot
\left(
\bm{\widehat{\beta}}
(y[1], y[2], \ldots y[N-1], y_{\text{out}})
-
 \right.
 \\&
 \left.
\bm{\widehat{\beta}}
(y[1], y[2], \ldots y[N-1])
\right)
,
\end{split}
\end{align}
where~$\bm{\widehat{\beta}}
(y[1], y[2], \ldots y[N-1])$
is the estimator
without outliers
and
$\bm{\widehat{\beta}}
(y[1], y[2], \ldots y[N-1], y_{\text{out}})$
is the estimator
contaminated with an outlier~$ y_{\text{out}}$.
For a better graphical analysis,
the SC results
are shown as the mean absolute value of~SC~(MASC)
of all estimators.
This unified measure was
proposed in~\cite{Bayer2020}
for multiparametric evaluations.

The MASC
considering $5\%$
of outliers
with their values ranging from $1$ to $20$
and
$1000$ Monte Carlo replications
is presented in Figure~\ref{f:masc},
showing that the WMLE is more robust to
outliers when compared to the MLE, for
all evaluated outlier values and signal lengths.
The MLE displays a MASC maximum value of about~$130$,
while the WMLE does not show values higher than~$5$.
Additionally,
the WMLEs present similar behavior
regardless of the evaluated
signal length and
outlier value,
whereas the MLEs display
higher MASC values as the signal length
and the outlier value increases.

In summary,
the Monte Carlo simulations
show that
the robust-based Rayleigh regression model parameter estimators
are not strongly influenced
by the presence or absence
of outliers in the observed signal.
Furthermore,
in real-world scenarios,
identifying
whether a signal is contaminated by outliers
is not an easy task.
Thus, using a robust approach to estimate
the parameters of the Rayleigh regression model
can avoid inaccurate inferences.

\section{SAR Image Study}
\label{s:appli}

In this section,
experiments
with two measured SAR data sets are presented
to demonstrate
the applicability of the proposed
approach
in SAR image analysis.
We employed the introduced
robust scheme
to detect
ground type
and anomalies
in SAR image scenes.
In particular,
VHF wavelength-resolution SAR images
are almost speckle-free,
since there might only be a single scatter
in the resolution cell~(at least for the CARABAS~II data set).
On the other hand, non-wavelength-resolution SAR images
are characterized by the possible presence
of more than one strong scatter in the resolution cell area,
which makes the speckle noise not negligible.
We emphasize that the speckle
and other random effects
are accommodated in the Rayleigh-distributed
output signal~$Y[n]$.
The regression structure models
the mean of~$Y[n]$ which is deterministically
affected by parameters and known covariates~(input).

\subsection{Ground Type Detector}
\label{s:gtd}

In this experiment,
we considered the
ground type detection
methodology used in~\cite{Palm2019}
to distinguish between three regions
in two SAR images extracted from
CARABAS~II and OrbiSAR
data sets
considering the proposed robust approach.

\subsubsection{CARABAS~II}

As reported in~\cite{Lundberg2006}
and~\cite{Ulander2005},
the CARABAS~II is a~VHF
wavelength-resolution system,
which means that the images have
almost no speckle noise.
The system operates with horizontal~(HH) polarization, and
the spatial resolution is~$2.5~\text{m}$ in both
azimuth and slant range.
The CARABAS~II images are
(i)~represented as
matrices of
$3000 \times 2000$
pixels
(each pixel size is~$1 \, \text{m} \times 1 \, \text{m}$),
corresponding
to
an area of $6~\text{km}^2$,
covering
a scene
of size
$3 \, \text{km} \times 2 \, \text{km}$;
(ii)~georeferenced
to the Swedish reference system RR92;
and
(iii)~available
in~\cite{data}.
The ground scene is
dominated by boreal forest with pine trees.
Fences, power lines, and
roads were also present in the scene.
Military vehicles~(targets)
were deployed in the SAR scene and
placed uniformly
in a manner to facilitate
their detection in the tests~\cite{Lundberg2006}.
The image
has~$25$ targets
of
three
different
sizes,
and
the spacing
between the vehicles is about 50 meters.

\begin{figure}
	\centering
	\includegraphics[scale=0.33]{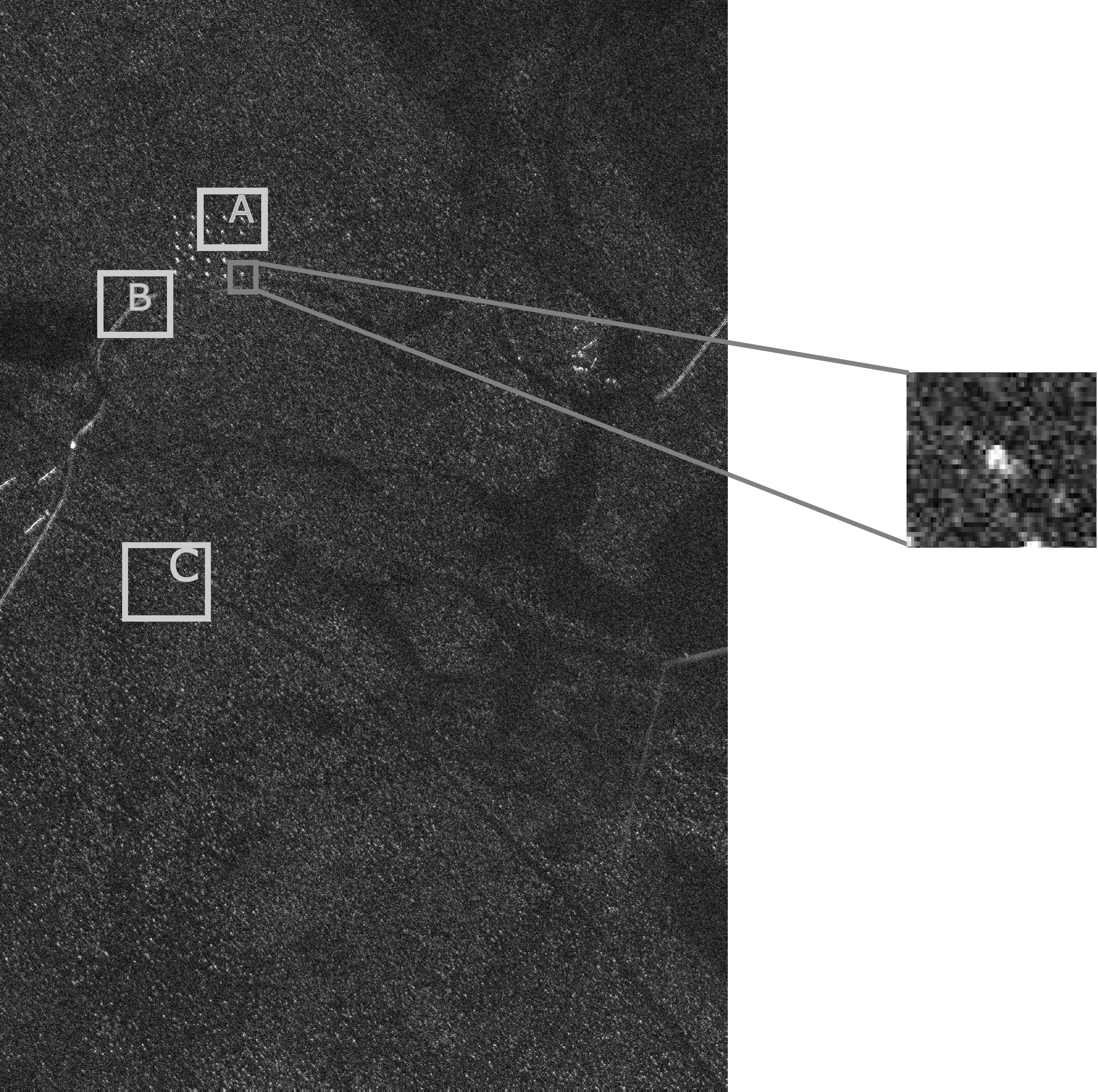}
	\caption{CARABAS~II
		image
		used in the regression models showing the tested regions.
		Regions A, B, and C
		represent
		areas
		containing military vehicles,
		power lines, and forest,
		respectively.}
	\label{f:data}
\end{figure}

In~\cite{Palm2019}, it was
computed the difference
in the behavior
among
the
lake, forest,
and military vehicles region.
The forest and
lake regions
in the CARABAS~II SAR image
characterize
most of the image area,
and
they
follow a homogeneous pattern.
The military vehicles
deployed in the SAR scene
introduce more
representative
behavior changing
when compared
to the forest and
lake regions~(homogeneous areas).
Additionally,
pixels
related to the power lines
show similar amplitude values
with the targets
and, consequently,
are strongly related
to the false alarm detection in this particular data set,
as discussed in~\cite{Lundberg2006}.
Furthermore,
as
both targets and power line structures
present
a different pattern from the rest of the image,
they may
be considered
as
anomalies observations~(outliers).
Thus, we adopted the detection methodology
proposed in~\cite{Palm2019}
to distinguish among
an area containing military vehicles,
power lines,
and forests---referred to as
Regions~A,~B, and~C,
respectively---,
which
are displayed in
Figure~\ref{f:data}.
For such,
we
modeled
the response signal mean
considering
an intercept, $x_1[n]=1, \,  \forall n $,
and
two dummy variables, $x_2[n]$~and~$x_3[n] $,
representing
each evaluated region.
The fitted model is
given by
\begin{align}
\label{e:modelequation}
g(\mu[n]) = \beta_1 +  \beta_2 x_2 [n]  + \beta_3 x_3 [n]
,
\end{align}
where
(i)~$y[n]$
is
the
vectorized
magnitude
pixels of
Regions~A,~B, and~C;
(ii)~variable~$x_2[n] = 1$,
for
Region~B, and zero for the rest;
(iii)~variable~$x_3[n] = 1$,
for Region~C, and
zero for the others;
and
(iv)~Region~A
is represented
for both~$x_2[n]$
and~$x_3[n]$ equal to zero.

To detect
the ground type,
the following hypotheses
are tested
\begin{align}
\label{e:detec}
\begin{cases}
\mathcal{H}_0 :
\beta_i= 0
,
\\
\mathcal{H}_1 :
\beta_i \neq 0
,
\end{cases}
\end{align}
for~$i = 2, 3, \ldots , k$.
The
evaluated ground types
are detected
when
the null hypothesis~\eqref{e:detec}
is rejected,
i.e.,~$T_W > \gamma$.
We
also obtained the detection results
based on
(i)~the Gaussian and Gamma regression models
considering a robust estimation process
and
(ii)~the
Rayleigh regression model based on
the maximum likelihood estimation
scheme
for comparative purposes.
To implement the detectors through
the
Gaussian- and Gamma-based
regression models considering a robust approach,
the {\tt R} function {\tt glmrob}~\cite{robustbase}
was used.
To perform the ground type detection,
the probability of false alarm was fixed to~$0.05$,
which is
a convenient
cutoff level to reject the null hypothesis~\cite{fisher1992},
and it is
widely employed
in signal detection applications~\cite{sevgi2009hypothesis,
	ru2016normalized,maleki2011energy,weber2006increasing}.
The fitted models
can be found in
Table~\ref{t:fit}.
Note that
the estimated values are not the same
as those
presented in~\cite{Palm2019},
since we evaluated
different regions
in the present experiment.
To perform the
robust estimation in the Rayleigh regression model,
we employed~$\delta=0.001$.
Considering
a probability of false alarm equal to~$0.05$,
the~$p$-values
of the Wald test
presented in Table~\ref{t:fit}
show
that all variables
in the proposed robust scheme
are significant,
i.e.,
the
null
hypothesis
in~\eqref{e:detec}
can be rejected,
and consequently,
a
correct detection of all evaluated land types
is indicated.
On the other hand,
the variable~$x_2[n]$
is not significant
for the other evaluated regression models,
i.e.,
Rayleigh regression model
based on non-robust estimation process
and the models using the
Gaussian- and Gamma-based
on a
robust estimation process
can not
distinguish the power line regions,
showing the importance
of robust methods
based on suitable distributions
to deal with outliers in SAR image modeling.

\begin{table}[t]
	\centering
	\caption{Fitted regression models for
		Regions~A,~B,
		and~C
	}
	\label{t:fit}
	\begin{tabular}{lccc}
		\toprule
		& Estimate & Standard Error &  $p$-value \\
		\midrule
		\multicolumn{4}{c}{Rayleigh regression model~(robust estimators)}  \\
		\midrule
		$\widehat{\beta}_1$ & $-1.6681$ &   $0.0333$ &    $< 0.001$   \\
		$\widehat{\beta}_2$ &  $0.1168$ & $0.0521$ &  $0.0250$ \\
		$\widehat{\beta}_3$ & $-0.4993$ & $0.0521$ & $< 0.001$  \\
		\midrule
		\multicolumn{4}{c}{Rayleigh regression model~(non-robust estimators)}  \\
		\midrule
		$\widehat{\beta}_1$ & $-1.4916$ & $0.0333$ &  $< 0.001$ \\
		$\widehat{\beta}_2$ &  $-0.0555$ &    $0.0521$ & $0.2867$   \\
		$\widehat{\beta}_3$ &  $-0.6759$ & $0.0521$ & $< 0.001$ \\
		\midrule
		\multicolumn{4}{c}{Gaussian regression model~(robust estimators)}  \\
		\midrule
		$\widehat{\beta}_1$ &  $0.1896$ &   $0.0090$ & $< 0.001$ \\
		$\widehat{\beta}_2$ & $0.0098$ &  $0.0140$ &    $0.4870$     \\
		$\widehat{\beta}_3$ & $-0.0738$ &   $0.0140$ & $< 0.001$ \\
		\midrule
		\multicolumn{4}{c}{Gamma regression model~(robust estimators)}  \\
		\midrule
		$\widehat{\beta}_1$ &  $5.5817$ &   $0.2558$ & $< 0.001$   \\
		$\widehat{\beta}_2$ &  $-0.4396$ &     $0.3823$  &    $0.2500$      \\
		$\widehat{\beta}_3$ & $2.6408$ &     $0.5209$ &  $< 0.001$  \\
		\bottomrule
	\end{tabular}
\end{table}

\subsubsection{OrbiSAR}

The SAR image
was acquired with the airborne OrbiSAR sensor of Bradar over
S\~ao Jos\'e dos Campos, Brazil.
As reported in~\cite{barreto2016classification}
and~\cite{shiroma2016dual},
the OrbiSAR is an airborne multipolarized
dual-band SAR
system
of Bradar~(formerly Orbisat), Brazil, that operates at
X-band with~HH polarization
and spatial resolution of~$1 \, \text{m}$,
and at P-band
with full polarization---HH,
vertical~(VV), VH, and HV---and spatial resolution of~$2 \, \text{m}$,
with three antennas mounted
on the same platform allowing repeat-pass and
multibaseline interferometry,
at~P- and~X-bands, respectively.
The system
is also
equipped with a state-of-the-art
navigation system and motion
compensation~\cite{shiroma2016dual}.
Additionally,
the bandwidth
and the radar swath can
be up to
400 MHz
and 14 km, respectively~\cite{shiroma2016dual}.

\begin{figure}
	\centering
	\includegraphics[scale=0.33]{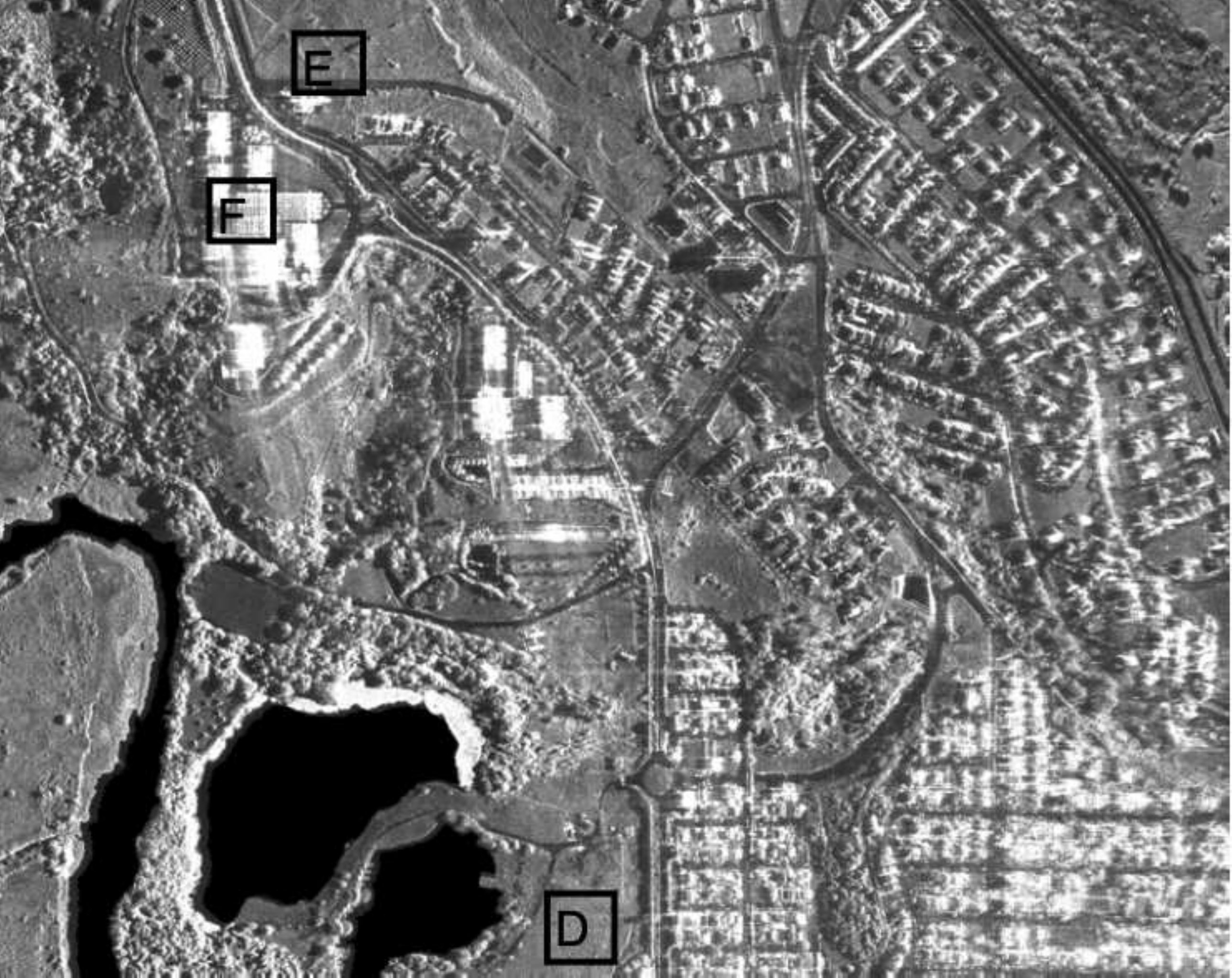}
	\caption{OrbiSAR
		image
		used in the regression models showing the tested regions.
		Regions D, E, and F
		represent
		an
		open area,
		a road, and an urban land type,
		respectively.}
	\label{f:bradar}
\end{figure}

Figure~\ref{f:bradar} shows an
X-band SAR image
acquired with the OrbiSAR system.
The image is represented in a~$2500 \times 3150$
matrix of magnitude data.
The ground scene of the
considered image
is dominated by
urban area
(light ground--top and bottom right area),
rivers (dark ground--bottom and left part of the image),
forests, roads, and open areas (gray
ground).
The urban area presents
a different pattern from the rest of the image
and
may
be considered
outliers.

To perform
the
ground type detection
in the X-band SAR image,
we adopted the same methodology
described
in the previous subsection
to distinguish among
an
open area,
a road, and an urban land type---referred to as
Regions~D,~E, and~F,
respectively.
Figure~\ref{f:bradar}
shows
the three different evaluated regions.
The
mean of the response signal was modeled using
the model presented in~\eqref{e:modelequation},
where
the response signal is composed of the
vectorized
magnitude
pixels of the
Regions~D,~E, and~F,
and~$x_2[n]$
and~$x_3[n]$ are dummy variables
related to regions~E and~F, respectively.

The fitted models
are shown in
Table~\ref{t:fitbradar}, with~$\delta=0.001$
for the proposed
method.
The $p$-values show that all
variables
in Rayleigh and Gamma regression models
considering a robust estimation process
are significant for a probability of false alarm equal to~$0.05$,
i.e.,
the
null
hypothesis
in~\eqref{e:detec}
can be rejected,
indicating a
correct detection of all evaluated ground types.
In contrast,
the variable~$x_2[n]$
is not significant
for the non-robust
Rayleigh regression method
and robust Gaussian approach,
i.e.,
these models
can not
distinguish the road region,
evidencing the importance
of suitable models
to deal with outliers in SAR image modeling.

\begin{table}[t]
	\centering
	\caption{Fitted regression models for
		Regions~D,~E,
		and~F
	}
	\label{t:fitbradar}
	\begin{tabular}{lccc}
		\toprule
		& Estimate & Standard Error &  $p$-value \\
		\midrule
		\multicolumn{4}{c}{Rayleigh regression model~(robust estimators)}  \\
		\midrule
		$\widehat{\beta}_1$ &   $-0.6069$   & $0.0423$ &  $< 0.001$ \\
		$\widehat{\beta}_2$ & $-0.1190$    & $0.0598$ & $0.0464$    \\
		$\widehat{\beta}_3$ & $2.7767$ & $0.1319$ &  $< 0.001$\\
		\midrule
		\multicolumn{4}{c}{Rayleigh regression model~(non-robust estimators)}  \\
		\midrule
		$\widehat{\beta}_1$ & $-0.6069$ & $0.0423$ &  $< 0.001$   \\
		$\widehat{\beta}_2$ & $-0.1105$ &  $0.0598$ &  $0.0646$     \\
		$\widehat{\beta}_3$ & $2.6749$ & $0.1319$ & $< 0.001$ \\
		\midrule
		\multicolumn{4}{c}{Gaussian regression model~(robust estimators)}  \\
		\midrule
		$\widehat{\beta}_1$ & $0.6087$ &  $0.1038$ & $< 0.001$  \\
		$\widehat{\beta}_2$ &  $-0.0777$ &  $0.1468$ &   $0.5970$        \\
		$\widehat{\beta}_3$ & $6.6209$ &  $0.3241$  & $< 0.001$  \\
		\midrule
		\multicolumn{4}{c}{Gamma regression model~(robust estimators)}  \\
		\midrule
		$\widehat{\beta}_1$ &  $1.6312$ &  $0.0338$ &  $< 0.001$ \\
		$\widehat{\beta}_2$ & $0.2542$ & $0.0517$ &  $< 0.001$ \\
		$\widehat{\beta}_3$ & $-1.5122$ & $0.0346$  &  $< 0.001$ \\
		\bottomrule
	\end{tabular}
\end{table}

\subsection{Anomaly Detection}

We propose a detection scheme
to detect anomalies in a SAR image
considering the Rayleigh regression model residuals.
Our methodology
aims
at detecting area changes,
measuring the deviations
of the observed pixel values~$y[n]$
from their estimated mean values~$\widehat{\mu}[n]$.
For such,
we use the quantile residuals~\cite{Dunn1996},
which are defined
as
\begin{align}
r[n]=\Phi^{-1}\left( F(y[n];\widehat{\mu}[n])\right)
,
\end{align}
where~$\Phi^{-1}$ denotes the standard normal
quantile function.
The quantile residual
can
detect
poor fitting
in regression models
and
follows an approximately standard Gaussian distribution~\cite{Dunn1996}
if the model is
correctly
specified.
If the estimated mean
of the response signal
is too far from the observed pixel value---which
can be highlighted by the residual values---then
an anomaly is detected.
To capture a model mismatch,
we adopt residual based control charts,
which
have been already used
in change detection in
remote sensing data,
e.g., in~\cite{Bayer2020} and~\cite{brooks2013}.
The introduced anomaly detection methodology
is based on the following
premises:
\begin{itemize}
	\item If the model is correctly fitted,
	then
	it is expected that the residuals are
	randomly distributed around zero
	and
	inside the interval~$[-3, \, 3]$,
	about~$99.7\%$
	of the observations~$
	\left(
	2\Phi (L) - 1
	\vert_{L=3}
	\approx
	99.7 \% \right)
	$.
	Consequently,
	the control limit $L$
	can be set equal to three~\cite{brooks2013,bayer2019}.
	\item
	If the residual value
	is outside the interval~$[-3, \, 3]$,
	then
	the analyzed pixel
	is understood
	to differ
	from the
	expected
	behavior
	according to the
	Rayleigh regression model
	fitted in
	the
	region
	of interest
	and,
	consequently,
	an anomaly
	is detected.

\end{itemize}

A post-processing
step using
mathematical
morphological operations,
such as
erosion,
dilation,
opening,
and
closing
operations,
can be considered
aiming at
(i)~removing
small
spurious pixel groups
which are regarded as noise
and
(ii)~preventing
the
splitting
of the interest objects
into multiple
substructures~\cite{gonzalez2008}.
The
anomaly detection method
used in the current  experiment
is summarized in
Algorithm~\ref{a:alg1}.

\begin{algorithm}
	\centering
	\caption{
		Anomaly detection method
		based on the robust
		Rayleigh regression method
	}
	\label{a:alg1}
	\begin{algorithmic}
		\REQUIRE
		Interest image~$\mathbf{X}_{\text{I}}$

		\ENSURE
		Detected results~$\mathbf{X}_{\text{D}} $
		\STATE 1) Select a region
		of interest~(training sample)~$\mathbf{X}_{\text{S}} \subset
		\mathbf{X}_{\text{I}}$.
		\STATE 2)
		Fit the robust Rayleigh regression method
		considering the~$\mathbf{X}_{\text{S}}$ image.
		\STATE 3)
		Using the fitted model obtained in 2),
		compute
		the
		residuals~$r[n]$
		of
		$\mathbf{X}_{\text{I}}$.
		\STATE 4)
		Obtain a binary images as follows:
		\IF{$(r[n]  \leq L) \,\,\, \text{or} \,\,\, (r[n]  \geq L)$}
		\STATE $ X^\star[n] \leftarrow 1 $
		\ELSE
		\STATE ${X}^\star[n] \leftarrow 0$.
		\ENDIF
		\STATE 5) Apply
		morphological operators as a final post-processing step:
		$\mathbf{X}_{\text{D}}
		\leftarrow
		\text{post-processing}(\mathbf{X}^\star)$.

	\end{algorithmic}
\end{algorithm}

To perform the proposed
detection method in a CARABAS~II SAR image,
we selected a region
containing military vehicles~(anomalies),
as shown in
the dark gray rectangle in Figure~\ref{f:data}.
This region
has about~$5\%$ of the
observations
related to
outliers.
According to the Monte Carlo simulations,
the WMLEs in the Rayleigh
regression model
are not strongly influenced
by the presence or absence
of outliers in the observed signal.
Additionally,
because in practical situations
it is difficult to identify
whether
there are outliers or not in the training sample,
we
selected
as~$\mathbf{X}_{\text{S}}$
a region with anomaly observations
to highlight the importance
of using a robust approach to deal with outliers.

To fit the regression model,
we considered as covariates
the other three images
with the same flight pass
available in the CARABAS~II
data set
to describe the amplitude mean value of
the CARABAS~II image pixels.
The model is specified for the
response signal mean
as follows
\begin{align}
g(\mu[n]) = \beta_1 +  \beta_2 x_2 [n]  + \beta_3 x_3 [n] + \beta_4 x_4 [n]
.
\end{align}
The response signal is composed of the
vectorized
amplitude values of
the training sample pixels.
Variables~$x_2 [n] $,~$x_3[n]$,
and~$x_4 [n] $
are the vectorized magnitude pixels
of
the images related to
pass one and
missions two, three, and four,
respectively.
As
we expect to have outliers in the
observed signal and not in the covariates,
$x_2 [n] $,~$x_3[n]$,
and~$x_4 [n] $
represent a forest area.
For the post-processing step,
we employed
an opening
operation---considering
a
$3 \times  3~\text{pixels}$
square structuring element,
whose size is linked by the
system resolution---followed by a
dilation---using
$7 \times  7~\text{pixels}$
structuring element
which is related
to the approximate size
of the military vehicles.
Using such operations,
we kept the criterion defined in~\cite{Lundberg2006},
i.e., detection with less than ten meters apart
are merged as one.

\begin{figure*}
	\centering
	\subfigure[Detections: Robust-based estimation]{
		\includegraphics[scale=0.4]{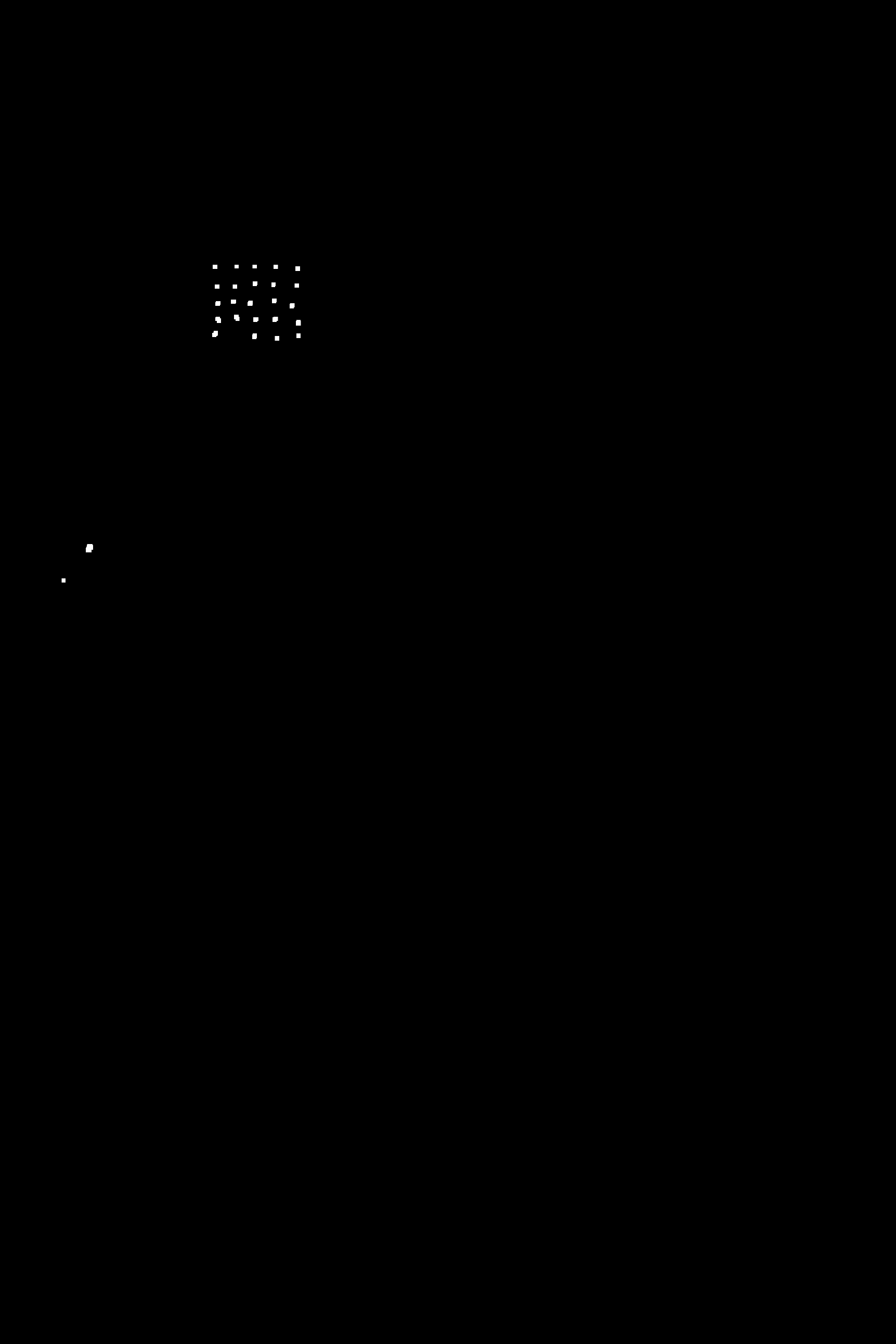}
		\label{f:dec1}}
	\subfigure[Detections: Non-robust-based estimation]{
		\includegraphics[scale=0.4]{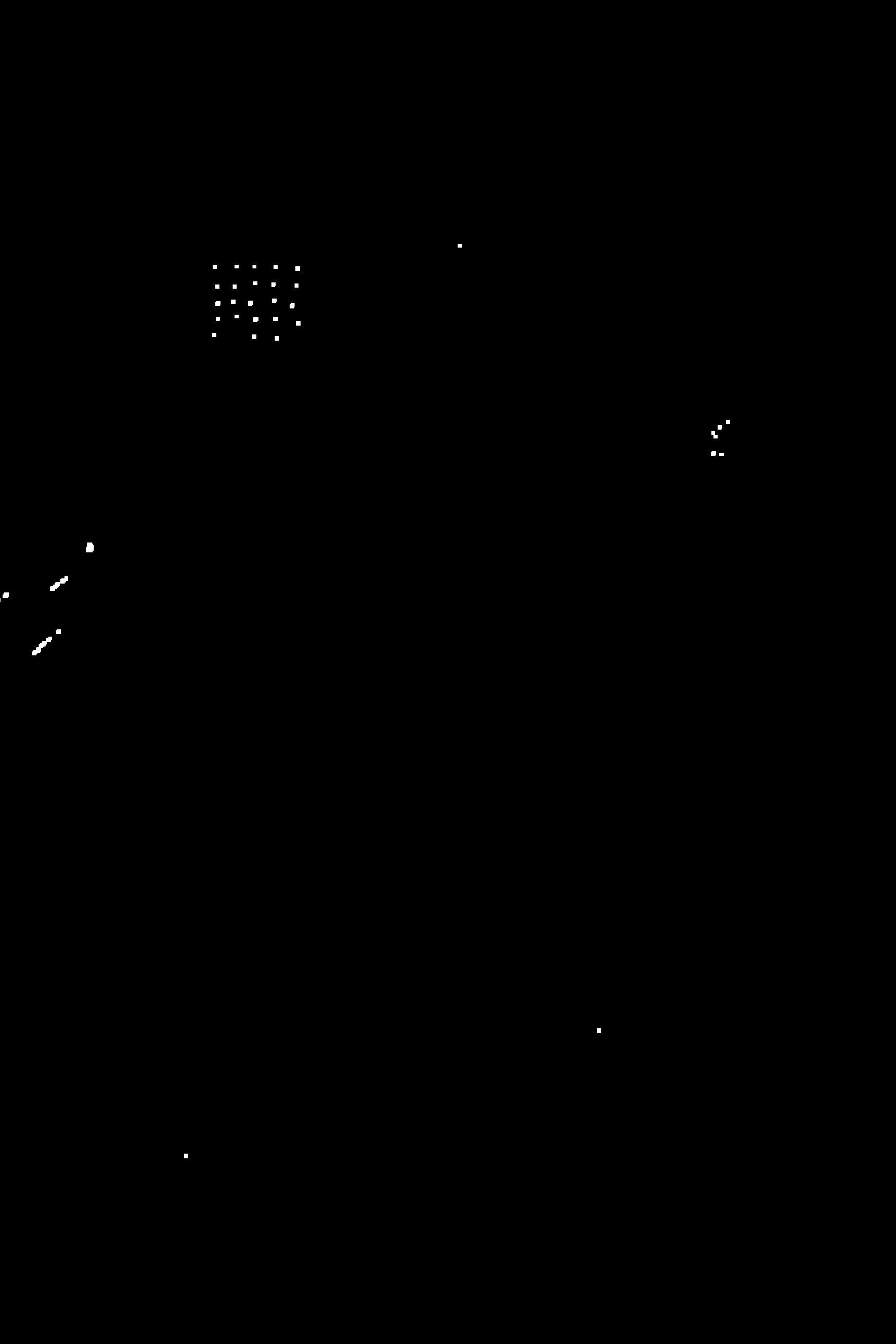}
		\label{f:dec2}}
	\caption{Detection results based on the
		robust and non-robust estimation of the Rayleigh regression model parameters.}
	\label{f:detec}
\end{figure*}

The anomaly detection results can be found
in Figure~\ref{f:detec}.
We compared the
detection
results
of
the
Rayleigh regression models
considering a robust and non-robust
estimation process,
which
are displayed in
Figure~\ref{f:detec}.
The robust method
detected~$24$ military vehicles
and two false alarms.
In contrast,
the non-robust scheme
can only
detect~$23$
military vehicles
and shows~$15$ false alarms.
In particular,
the false alarms
in the non-robust method
are related to the power lines area;
this result is in accordance with the ones presented in
the Section~\ref{s:gtd},
showing that,
in both experiments,
the non-robust estimation process
can not distinguish between the targets and
power line areas,
and consequently,
evidencing the importance
of a robust approach
to deal with outliers.

We also compared the proposed methodology
with three
different approaches
presented
in~\cite{Palm2020},~\cite{Lundberg2006}, and~\cite{Vu2018};
the performance of the
proposed
scheme
was very close to the competing methods
specifically developed to this aim:
only one less detection hit;
and two more false alarms
than the results showed
in~\cite{Palm2020} and~\cite{Vu2018}.
On the one hand, the accurate performance
of such methods
is related to a
optimized
threshold choice.
On the other hand,
our proposed anomaly detection method
shows accurate detection results
avoiding this step,
since
the residual-based
control chart
has a fixed
theoretical
threshold~($L=3$).
The detection results
of all evaluated methods
are summarized
in Table~\ref{t:comp}.

\begin{table}
	\centering
	\caption{
		Number of Detected Targets and False Alarms
		obtained considering the
		Rayleigh regression models
		based on robust and non-robust
		approaches, and the methods in~\cite{Palm2020},~\cite{Lundberg2006},
		and~\cite{Vu2018}}
\label{t:comp}
	\begin{tabular}{cccc}
		\toprule
		Method & Detected Targets & False Alarm \\
		\midrule
		Robust Approach & $24$ & $2$ \\
		Non-robust Approach & $24$ & $15$ \\
		Method in~\cite{Palm2020} & $25$ & $0$ \\
		Method in~\cite{Lundberg2006} & $25$ &  $2$\\
		Method in~\cite{Vu2018} & $25$ & $0$\\
		\bottomrule
		\end{tabular}
\end{table}

\section{Conclusions}
\label{s:con}

This paper introduced
robust estimators
for the Rayleigh regression model parameters.
In particular,
we employed
the
weighted maximum likelihood approach
to obtain
estimators that are
robust to the presence of outliers.
Monte Carlo simulation results showed that
the WMLEs
outperformed traditional MLEs
in terms
of
relative bias and
root mean square
error.
In particular,
the non-robust estimators
presented
a relative bias value
$65$-fold larger than the results provided by the
robust estimators
in signals corrupted with outliers.
In terms of
sensitivity analysis and break down point,
the robust approach
resulted in a reduction
of about~$96\%$
and~$10\%$ in
the mean absolute value
in compassion to the non-robust estimators.
For non-contaminated signals,
both schemes
had similar behavior.
Two studies
considering
the proposed
robust approach
in the Rayleigh regression model
parameter estimation
to distinguish between different regions in a SAR image
were presented and discussed,
showing competitive detection results
compared to the non-robust Rayleigh-,
robust Gaussian-, and robust Gamma-based measurements.
Moreover,
we proposed an anomaly detector based
on the Rayleigh regression model.
The robust estimation approach
excelled  in terms of detection
when compared to
the non-robust estimators of the
Rayleigh regression model parameters,
and
it offered
very close results
to those reported
in~\cite{Palm2020},~\cite{Lundberg2006},
and~\cite{Vu2018},
i.e.,
only one less detection hit;
and two more false alarms
than the results showed
in~\cite{Palm2020}
and~\cite{Vu2018}.

\section*{Acknowledgements}

This work was supported in part by
Conselho Nacional de Desenvolvimento Cient\'ifico e Tecnol\'ogico~(CNPq),
Coordena\c{c}\~ao de Aperfei\c{c}oamento de
Pessoal de N\'ivel Superior~(CAPES),
Project Pr\'o-Defesa IV, Brazil,
Swedish-Brazilian Research
and Innovation Centre~(CISB), and Saab AB.

{\small
\singlespacing
\bibliographystyle{siam}
\bibliography{rayleigh}
}

\end{document}